\documentstyle[preprint,revtex]{aps}
\begin{document}
\draft
\def\tr{{\rm tr}}
\def\Tr{{\rm Tr}}
\def\dim{{\rm dim}}
\def\SO{{\rm SO}}
\def\SU{{\rm SU}}
\def\E{{\rm E}}
\def\U{{\rm U}}
\def\half{{\textstyle{1\over2}}}
\hyphenation{anom-aly anom-alies coun-ter-term coun-ter-terms}
\preprint{IFP-440-UNC}
\preprint{hep-th/9210049}
\preprint{September 1992}
\begin{title}
Heterotic Parafermionic Superstring
\end{title}

\author{Paul H.~Frampton and James T.~Liu}
\begin{instit}
Institute of Field Physics, Department of Physics and Astronomy, \\
University of North Carolina, Chapel Hill, NC 27599-3255
\end{instit}

\begin{abstract}
\hskip-\parindent\hskip\parfillskip{Abstract}\par
\smallskip
Superstrings have been postulated based on
parafermionic partition functions which
permit spacetime supersymmetry by generalized
Jacobi identities.  A comprehensive search finds new
such identities. Quadrilateral anomaly cancellation
gives constraints on allowed chiral fermions.  Bosonic left-movers and
$Z_4$ parafermionic right-movers combine in a new heterotic superstring,
more constrained than the old one, yet equally applicable to
physics.
\end{abstract}

\narrowtext

Progress in superstring theory has been slow during the last few years
compared to its great strides about eight years ago.  Part of the problem
is that the most promising string theory, the heterotic string, has too much
non-uniqueness in its low-energy predictions.  This stems from the lack of
calculational methods for selecting the correct ground state, for breaking
supersymmetry and for explaining the vanishing cosmological constant.  None
of these deep problems will be addressed here, but we adopt the viewpoint
that the heterotic string {\it does} have too much freedom and that one
should seek a more restrictive starting point.

Parafermionic strings have been considered for several years\cite{FU} and
have recently been intensively studied in an important series of
papers by the Cornell
group\cite{CI,CII,CIII,CIV}.  The motivation is to reduce the
critical spacetime dimension below $d_c=10$ by increasing the symmetry
of the worldsheet.  In the $d_c=10$ superstring
there is worldsheet supersymmetry which pairs a boson with a fermion
for each dimension; in a parafermionic superstring each boson is paired
with a $Z_K$ parafermion thus realizing a worldsheet
{\it fractional} supersymmetry\cite{CI}.

In the general case, the left-movers and right-movers of a
parafermionic string are assumed to carry $Z_K$ parafermionic fields of
order $K_L$ and $K_R$ respectively, denoted $(K_L,K_R)$.
%
%
The critical dimension for this theory is found to be
$d_c=2+16/{\rm max}(K_L,K_R)$ except for the special case $(1,1)$ which
has $d_c=26$\cite{FU,CI}.  The (1,1) bosonic string which is the
simplest and oldest string theory has inadequacies such as a tachyon,
no fermions and no finiteness.  Extending this model to a (2,2)
superstring solves all these problems, but this Type II model has
insufficient freedom to accommodate the low-energy physics of the
Standard Model\cite{DKV}.  The (1,2) heterotic superstring has, on the other
hand, perhaps too much freedom with respect to low-energy predictions.

The hope of parafermionic strings is to provide a more restrictive
scenario for string model-building.  For $d_c\ge4$ there are seven new
parafermionic models, three with $K_R=4$ ($K_L=1,2,4$) and four with
$K_R=8$ ($K_L=1,2,4,8$).
In this
article we shall examine these new models from the points of view
of spacetime consistency (anomalies) and
of possible connections to physics.
Without knowledge of the fractional
superconformal constraint algebra\cite{CIV},
we work only at the partition function level.

{\it Six-dimensional $K=4$ models:}
Models with $K_L\le K_R=4$ have critical dimension $d_c=6$.  In analogy with
the Type II string, the (4,4) model is the most constrained of these three.
The partition function for the (4,4) string,
$Z_{(4,4)}$, is composed of $Z_4$ parafermionic string functions and
was first written down in Ref.~\cite{CI}.  It is unique
under the assumption that the individual components
are tachyon free ({\it i.e.}\ have
a series expansion in non-negative powers of $q$).

Modular invariance itself does not fix the normalization of $Z_{(4,4)}$,
denoted
here by $\alpha$,
but a physical interpretation of the partition function as
counting states requires integer multiplicities, {\it i.e.}\ $16\alpha\in Z$.
The leading expansion of $Z_{(4,4)}$ gives
$Z_{(4,4)}=\alpha(4-4)^2q^0+\ldots$ which is identified as four bosons and four
fermions coming from each side of the string with total multiplicity
$\alpha$.  The standard interpretation of these states being a vector
and spinor in the $d_c-2=4$ dimensional transverse space (thus forming
a super-Maxwell multiplet) requires the stronger condition
$\alpha\in Z$.

Although the massless states can be identified as having usual bosonic and
fermionic nature, the full spectrum contains states at mass level 0 and
$\half$ (mod 1) of which the latter
have well-known\cite{CII} difficulties with spin and statistics.
Here we focus on the massless spectrum
and assume the low-energy analysis is unaffected by the resolution of such
difficulties encountered at the massive level.

With $\alpha=1$, the massless states are created by tensoring $d=6$
super-Maxwell multiplets on the left and right giving $N=4\rm A$ or 4B
six-dimensional supergravity (in terms of $d=6$ symplectic
Majorana-Weyl spinors).  The $N=4\rm A$ theory is non-chiral because
the spinors on the left and right have opposite chirality; $N=4\rm B$
is chiral and is prone to spacetime anomalies\cite{AGW}.

The massless Type~4B string states form a $N=4\rm B$ graviton multiplet and
tensor multiplet and
the pure gravitational anomaly arising from this combination is
non-vanishing\cite{CIV,SGref4}.  In Ref.~\cite{CIV} it is
proposed to cancel the anomaly by
the judicious addition of twenty extra tensor multiplets.
At the partition function level, these additional massless states can
be accommodated by choosing the normalization $\alpha=6$.
However, since it is not clear how these new physical states
arise, we are lead to consider the other possibility that they come from
extra modular-invariant terms in $Z_{(4,4)}$ beyond those found in
\cite{CI}; we shall examine this below.

Now consider the generalized
heterotic (1,4) and (2,4) models.  Similarly to the (1,2) heterotic
string, the (1,4) model has a gauge group realized by a $c=26-d_c=20$
Kac-Moody algebra generated by the internal left-moving bosons.
Massless states in this model arise from tensoring the left-moving
spacetime vector and Kac-Moody currents with the super-Maxwell
multiplet of right-movers to give six-dimensional $N=2$ supergravity
coupled to super Yang-Mills theory.

In a general $N=2$ theory the possible multiplets are\cite{SGref4,SGref2}
\begin{eqnarray}
(a)\qquad&&(e_\mu^\alpha,\psi_{L\mu}^A,B_{\mu\nu}^{(-)}),\nonumber\\
(b)\qquad&&(B_{\mu\nu}^{(+)},\lambda_R^A,\phi),\nonumber\\
(c)\qquad&&(A_\mu^a,\chi_L^{aA}),\nonumber\\
(d)\qquad&&(\lambda_R^j,\phi^i)\qquad i=1\ldots4n_d,\ j=1\ldots2n_d,
\label{SGSYMtwo}
\end{eqnarray}
where $\mu,\alpha=1,2,3,4$ are transverse space indices and $A$ is in the 2
of Sp(2).
The leading gravitational and gauge anomalies have the forms
\begin{eqnarray}
I_8(R)&=&{1\over5760}(273n_a-29n_b+n_c-n_d)\tr R^4+\ldots,\nonumber\\
I_8(F)&=&{1\over24}(\Tr F^4({\rm adj})-\Tr F^4({\rm matter})),
\label{anom}
\end{eqnarray}
where the adjoint comes from the gauginos $(c)$ and the matter comes
from the gauge representations of $(d)$.  Since there is only one
graviton, $n_a=1$.  For the minimal $N=2$ matter content of the (1,4) model,
we find
that $n_b=1$, $n_c=\dim\;G$ and $n_d=0$ where $G$ is the gauge group.
The gravitational anomaly is thus proportional to $(244+\dim\;G)$ and
is hence non-zero for any $G$ of positive dimension.

Examining (\ref{anom}), we see that leading gravitational anomaly cancellation
requires the addition of either new tensor matter $(b)$ or hypermatter
$(d)$ multiplets.  However, tensor matter multiplets cannot be created
by tensoring bosonic left-movers with any right-moving states.  As a
result, we must add $n_d$ hypermatter multiplets with
$n_d=(244+\dim\;G)$ and in representations of $G$ such that the leading gauge
anomaly also vanishes.

Maximal gauge symmetry for the (1,4) model in $d_c=6$ arises for
$G=\SO(40)$ and requires $n_d=1024=2^{10}$ to cancel the $\tr R^4$
anomaly.  These hypermatter states must further be put into
representations of SO(40) so as to cancel the leading gauge anomaly.
The smallest dimensional irreps of SO(40) are 1,~40,~780,~819,$\ldots$
with leading quadrilateral
anomalies 0,~1,~32,~48,$\ldots$ respectively\cite{quadref}.
We find that anomaly cancellation requires
$n_{780}=1$, $n_1=244$, $n_{40}=n_{819}=0$ but
this leads
to a non-chiral model in $d=6$ since the hypermatter spinors can pair with
the gauginos.

One {\it can} obtain chiral examples in $d=6$ with other choices of
gauge group; as examples, we consider $\SO(24)\times \SO(16)$ and
$\SO(24)\times \E_8$ where the partition functions have been written in
Ref.~\cite{CII}.  For the $\SO(24)\times \SO(16)$ model, SO(24) has irreps
1,~24,~276,~299,$\ldots$ with quadrilateral anomalies respectively
0,~1,~16,~32,$\ldots$.  SO(16) carries irreps 1,~16,~120,~135,$\ldots$
with anomalies 0,~1,~8,~24,$\ldots$.
Leading gravitational and gauge anomaly cancellation
gives three conditions on the additional hypermatter representations,
and we find the only possible irreps of the hypermatter that cancel the
leading anomalies are (under (SO(24),SO(16)))
\widetext
\begin{eqnarray}
(276,1)+(1,120)+244(1,1)\qquad&&\hbox{non-chiral,}\nonumber\\
(276,1)+8(1,16)+236(1,1)\qquad&&\hbox{chiral under SO(16),}\nonumber\\
(1,120)+16(24,1)+136(1,1)\qquad&&\hbox{chiral under SO(24),}\nonumber\\
16(24,1)+8(1,16)+128(1,1)\qquad&&
\hbox{chiral under {\it both} SO(16) {\it and} SO(24).}
\label{acanII}
\end{eqnarray}
\narrowtext
Cancellation of the non-leading and mixed anomalies will require a
Green-Schwarz mechanism\cite{GS}.

For $G=\SO(24)\times \E_8$, a similar analysis shows that
the irreps of the hypermatter which cancel leading anomalies are
(under (SO(24),$\E_8$))
\widetext
\begin{eqnarray}
(276,1)+(1,248)+244(1,1)\qquad&&\hbox{non-chiral,}\nonumber\\
16(24,1)+(1,248)+136(1,1)\qquad&&\hbox{chiral under SO(24),}\nonumber\\
(276,1)+492(1,1)\qquad&&\hbox{chiral under $\E_8$,}\nonumber\\
16(24,1)+384(1,1)\qquad&&\hbox{chiral under {\it both} SO(24) {\it and}
$\E_8$.}
\label{acanIII}
\end{eqnarray}
\narrowtext

Since the SO(40), $\SO(24)\times \SO(16)$ and $\SO(24)\times
\E_8$ partition functions given in \cite{CII} do not contain additional
hypermatter multiplets, the chiral representations
of (\ref{acanII}) and (\ref{acanIII}) must arise from non-adjoint $G$ states
coming in a novel way from the left-moving bosonic string.
How these states may arise needs to be addressed if
we wish to more fully understand six-dimensional anomaly cancellation.

The remaining six-dimensional parafermion model is the (2,4) one where
the left-movers are those of a superstring compactified from $d=10$ to
$d_c=6$ while the right-movers are the $K=4$ set already considered.
The compactification of the left-movers may be toroidal or more
general.  Compactification on a torus will give a $N=6$
supergravity theory which is anomalous.

A more general compactification of the left-movers using {\it e.g.}\ a
free-fermion inter\-pretation\cite{bdg,abk,klt} can realize a $\hat
c=10-d_c=4$ super Kac-Moody symmetry with gauge group $G$ of dimension
up to 12 (in $d_c=6$).  When this non-abelian symmetry is
realized on the left movers, no massless fermions arise from the
left\cite{DKV}.  Thus the massless states form a $N=2$ theory similar to that
of the (1,4) model.  As before, the minimal partition function
will have a leading gravitational anomaly
which is non-vanishing but
can be cancelled by the addition of further massless
states.

When considering the (4,4) model, we alluded to the
necessity for additional massless states beyond those accommodated in
the $\alpha=1$ partition function of \cite{CI}.  We have therefore
considered the use of $Z_4$ parafermionic string functions $c^l_n$ with
odd $n$
(the string functions are related to the
parafermionic characters by $\chi^l_n(q)=\eta(q)c^l_n(q)$; see for
example \cite{CII,ZF,GQ}).
These odd functions did
not play a role in the analysis of Ref.~\cite{CII}.  An exhaustive search
for all tachyon-free modular-invariant (4,4) partition functions
leads to 15 independent choices of which 4 are vanishing and may be
spacetime supersymmetric.  One of these four is just $Z_{(4,4)}$,
and the other three can be written as
\widetext
\begin{eqnarray}
Z_4^{(2)}&=&(32|C_4^I|^2+|D_4^I|^2+32|E_4^I|^2)
-(32|C_4^{II}|^2+|D_4^{II}|^2+32|E_4^{II}|^2),\nonumber\\
Z_4^{(3)}=\overline{Z_4^{(4)}}&=&(32C_4^I\overline{C_4^{II}}
+D_4^I\overline{D_4^{II}}
+32E_4^I\overline{E_4^{II}})
-(32|C_4^{II}|^2+|D_4^{II}|^2+32|E_4^{II}|^2),
\label{newZ}
\end{eqnarray}
\narrowtext
where $C_4^I$, $D_4^I$ and $E_4^I$ are combinations of the even
string functions
\begin{eqnarray}
C_4^I&=&2(c^2_0)^3c^4_2+(d^{0+}_0)^3c^4_2+3d^{0+}_0(c^2_0)^2c^2_2,\nonumber\\
D_4^I&=&(d^{0+}_0)^4+8d^{0+}_0(c^2_0)^3-16(c^4_2)^4-16c^4_2(c^2_2)^3,\nonumber\\
E_4^I&=&4d^{0+}_0(c^4_2)^3+6c^2_0c^4_2(c^2_2)^2+d^{0+}_0(c^2_2)^3,
\label{CDE}
\end{eqnarray}
and $C_4^{II}$, $D_4^{II}$ and $E_4^{II}$ are combinations of the
odd ones
\begin{eqnarray}
C_4^{II}&=&4(c^1_1)^3c^3_1+4c^1_1(c^3_1)^3,\nonumber\\
D_4^{II}&=&(d^{0-}_0)^4,\nonumber\\
E_4^{II}&=&(c^1_1)^4+6(c^1_1)^2(c^3_1)^2+(c^3_1)^4,
\label{CDEII}
\end{eqnarray}
where $d^{0\pm}_0=c^0_0\pm c^4_0$.

Because these three new terms in (\ref{newZ}) are modular invariant by
themselves, they can be consistently added to the original (4,4)
partition function, $Z_{(4,4)}$.  However, by examining the parafermionic
string
functions, we see that $C_4$, $D_4$ and $E_4$ only have states at mass
level $5\over12$, $2\over3$ and $11\over12$ (all mod 1) respectively.
They hence contribute only massive states to the spectrum and have no
effect on the low energy properties of the (4,4) string.

The vanishing of the modular invariants, (\ref{newZ}), arises from the
new generalized
Jacobi identities relating even and odd string functions
\begin{equation}
C_4^I=C_4^{II},\qquad D_4^I=D_4^{II},\qquad E_4^I=E_4^{II},
\label{newid}
\end{equation}
which we have proven on the basis of modular invariant function theory
following the procedure given in \cite{CIII}.

We remark that the minus signs in (\ref{newZ}) may be interpreted either as
the statistics factor for fermions or as an internal GSO like
projection.  Without proper identification of spin and statistics, it
is not possible to make this distinction.
However, this issue is important since a GSO
projection actually {\it removes} physical states from the spectrum
whereas the other case does not.  Since the minus signs serve
to project out the tachyons (present in $D_4$ and $E_4$),
a natural interpretation is to view the
signs as a GSO projection.  With this interpretation, there are no
additional physical states arising from (\ref{newZ}).

{\it Four-dimensional $K=4$ models:}
In compactifying the $K=4$ sector from $d_c=6$ to the physical
spacetime $d=4$, since we have no complete understanding of the
underlying worldsheet CFT, it is impossible to make
categoric statements about compactifications without any geometrical
interpretation.  This is a {\it caveat} of our analysis, but we feel
confident that
the rank of the $d=4$ gauge group cannot exceed that allowed in
geometric compactification, namely
$r_{max}=(d_c-4)$.  Thus (4,4) compactified to $d=4$ can have an
internal non-abelian gauge group only with rank $r\le2$.

The (1,4) model in $d_c=6$ can have chiral fermions
transforming under a rank $r=20$ gauge group ({\it e.g.}\ $\SO(24)\times
\SO(16)$ or $\SO(24)\times \E_8$ {\it ut supra}), and this will
lead to an internal group with $r\le22$ in $d=4$.  For example, the
maximally symmetric case of (1,4) in $d=4$ is for gauge group SO(44)
and is a non-chiral $N=2$ model with partition function easily
constructed following the procedure given in \cite{CII}.
In order to obtain a chiral model, one needs to break at least one of these
supersymmetries to give $N\le1$ in $d=4$ dimensions.

For (2,4) models in $d=4$, because the left-movers are governed by a
SCFT, we can take the approach of Dixon {\it
et al.}\cite{DKV}.  Since there are only two internal dimensions for the
$K_R=4$ right-movers, at least some of the rank 4 standard model gauge
group would need to arise from the left-movers.  As long as a
non-abelian symmetry is realized, it then follows that no massless
fermions can arise from the left-movers\cite{DKV}.  Thus all massless
four-dimensional states take the form of bosonic left-moving states
tensored with (spacetime) supersymmetric right-movers.  If we assume
that right-moving fermions are uncharged under right-moving
symmetries, then any massless fermions must be right-movers which are
either gauge singlets or tensor producted with non-abelian generators
on the left.  As a result, we demand the entire gauge group to be
realized by a $\hat c=6$ super Kac-Moody algebra on the left.

We can now apply the result of \cite{DKV} that
if we seek $\SU(3)\times \SU(2)\times \U(1)$ with a realistic fermion
spectrum then we
need $\hat c\ge4+5/3+1=20/3$
which is too large to be accommodated by a $\hat c=6$ SCFT of the left movers.
This argument is not as rigorous as the one used in \cite{DKV} for the (2,2)
Type~II superstring because we have assumed that none of the gauge
group arises from the $K_R=4$ right-movers.

{\it Four-dimensional $K=8$ models:}
For $K=8$, the critical dimension is $d_c=4$ so that no
compactification is necessary (or allowed).  This is, however, a mixed
blessing because the
right-movers for all $K_R=8$ models are in $N=1$ supermultiplets
with fermion helicities $\pm\half$ so that the tensor product with an
independent $K_L=1,2,4$, or 8 sector will give non-chiral fermions.
To achieve chirality would require a correlation between left and right movers
which appears difficult without compactification.

In Ref.~\cite{CI}, one example of a tachyon-free modular-invariant
spacetime supersymmetric (8,8) partition function is provided.  We have
confirmed that this partition function is unique by an exhaustive
search of modular invariant combinations of all $K=8$ parafermion
characters.  Out of a total of 12 modular invariants, only 2
appropriate linear combinations are tachyon-free.  However, the second
tachyon free combination is non-supersymmetric so in this sense the
(8,8) partition function is unique and, unlike the $K=4$ case above, we find
no new $K=8$ identities beyond those of \cite{CIII}.

{\it Possible approaches to physics:}
Given the discussions of the present article, it is
clear that of all the new models the heterotic (1,4) model is the most
promising for accommodating the low-energy physics of the standard model.
For geometrically-interpretable compactification on $(M^4\times K^2)$,
however, there
is an insuperable hurdle for four dimensional chirality.  Although for higher
dimensions
there exist Ricci-flat manifolds which can preserve spacetime supersymmetry,
no such manifold exists for $K^2$
except a torus which leads to a non-chiral $N=2$ model.
Hence we require
a {\it non-geometric} reduction to $d=4$ from the (1,4) string in $d_c=6$.
This is precisely what has been suggested based on the independent
consideration of spin
and statistics by the Cornell group\cite{CII,CIV}.  The (1,4)
parafermionic heterotic superstring, quite unlike the less constrained and
much more familiar (1,2) heterotic superstring, may thus exist consistently
in $d=4$ {\it only}, and not in the critical dimension; chiral fermions must
then arise from $N=1$ right movers in a complex representation of the gauge
group generated by the left-movers.

The status of the mathematical consistency of
the (1,4) parafermionic superstring
is yet to be understood
at the level of the older (1,2) heterotic or
(2,2) Type~II superstring.
Construction of the worldsheet constraint algebra for the $K=4$
parafermionic superstring is thus an interesting issue and merits further
study.

\bigskip
This work was supported in part by the U.S.~Department of Energy under Grant
No.~DE-FG05-85ER-40219.

\end{document}